\documentclass[conference]{IEEEtran}
\IEEEoverridecommandlockouts
\usepackage{cite}
\usepackage{amsmath,amssymb,amsfonts}
\usepackage{algorithmic}
\usepackage{graphicx}
\usepackage{textcomp}
\usepackage{xcolor}
\usepackage[hidelinks]{hyperref}

\newcommand{\subsubsubsection}[1]{\paragraph{\textbf{#1}}}

\def\BibTeX{{\rm B\kern-.05em{\sc i\kern-.025em b}\kern-.08em
    T\kern-.1667em\lower.7ex\hbox{E}\kern-.125emX}}
\begin{document}

\title{{An Explorative Study on}\\Distributed Computing Techniques in Training and Inference of Large Language Models}

\author{
\IEEEauthorblockN{Sheikh Azizul Hakim\textsuperscript{†}}
\IEEEauthorblockA{\textit{Computer Science and Engineering} \\
\textit{Bangladesh University of Engineering and Technology}\\}
\and
\IEEEauthorblockN{Saem Hasan\textsuperscript{†}}
\IEEEauthorblockA{\textit{Computer Science and Engineering} \\
\textit{Bangladesh University of Engineering and Technology}\\}
\thanks{\textsuperscript{†}These authors contributed equally to this work.}
}
\maketitle

\begin{abstract}
\nocite{chatgpt}

Large language models (LLM) are advanced AI systems trained on extensive textual data, leveraging deep learning techniques to understand and generate human-like language. Today's LLMs with billions of parameters are so huge that hardly any single computing node can train, fine-tune, or infer from them. Therefore, several distributed computing techniques are being introduced in the literature to properly utilize LLMs. We have explored the application of distributed computing techniques in LLMs from two angles. 

\begin{itemize}
    \item We study the techniques that democratize the LLM, that is, how large models can be run on consumer-grade computers. Here, we also implement a novel metaheuristics-based modification to an existing system. 
    \item We perform a comparative study on three state-of-the-art LLM serving techniques. 
\end{itemize}
\end{abstract}

\section{Background}

%We compare various techniques introduced in the literature for LLM training and inference ... then we show an update in petals 

\subsection{Distributed Computing}
Distributed computing refers to a system where processing and data storage are distributed across multiple devices or systems rather than being handled by a single central device. In a distributed system, each device or system has its own processing capabilities and may also store and manage its own data. These devices or systems work together to perform tasks and share resources, with no single device serving as the central hub. The key characteristics of distributed computing include the following: Parallel Processing, Coordination, Scalability, Resource sharing, Fault tolerance, etc.

\subsection{Distributed and other High Performance Computing Aspects of ML}

In recent years, the fast development of new technologies has resulted in an unparalleled increase in data collection. Machine Learning (ML) methods are increasingly being utilized to evaluate datasets and construct decision-making systems where an algorithmic solution is not possible due to the problem's complexity. Examples include controlling self-driving cars \cite{bojarski2016end}, recognizing speech \cite{amodei2016deep}, or predicting consumer behavior \cite{khandani2010consumer}. The training data required for these sophisticated applications can easily be in the order of terabytes \cite{canini2012sibyl}.

Machine learning solutions are impacted by both data characteristics and the effectiveness of learning algorithms. Currently, a challenge exists in the incapacity of learning algorithms to efficiently utilize all available data within a reasonable timeframe for the learning process. Nevertheless, a significant portion of machine learning computations revolves around fundamental operations on vectors, matrices, or tensors—problems well-established in linear algebra. Addressing the optimization of such operations has been a focal point in high-performance computing (HPC) research for many years.

In recent years, deep learning has had significant success, particularly in domains like image processing, natural language processing (NLP), and speech recognition. The success can be attributed in part to the abundance of data and the growing size of deep learning models. As advancements in deep learning persist, enhancing the scalability, privacy, and interpretability of Distributed Neural Networks (DNN) becomes increasingly crucial.

\subsection{Large Language Models}
Large Language Models (LLMs) represent a groundbreaking advancement in the field of artificial intelligence, particularly in natural language processing. These models, often built on transformer \cite{vaswani2017attention} architectures, have the capacity to understand and generate human-like text with remarkable proficiency.  %They're massive, with lots and lots of parameters (think of them as settings) in the billions, which helps them pick up on the tiny details and rules of language from really big sets of data. 
These models are trained through a lot of practice on different kinds of text, making them know a whole bunch about language. They can do lots of interesting things like finishing sentences, translating languages, answering questions, and summarizing text. One famous example is ChatGPT, a version of GPT-3.5 refined with RLHF (reinforcement learning with human feedback). It is a powerful and widely used language model created by OpenAI. A few more examples of Large Language Models are GPT-3 \cite{floridi2020gpt}, BERT \cite{devlin2018bert}, T5 \cite{raffel2020exploring}, RoBERTa \cite{liu2019roberta} etc. Training LLMs with billions of parameters pose significant challenges and require considerable computational resources. For example, training a one-trillion-parameter GPT style model on 20 trillion tokens requires a staggering 120 million exaflops\footnote{An exaflop is one quintillion floating point operations per second.} of computation.

\subsection{Distributed Computing Techniques in Training and Inference of LLM}
The NLP community has discovered in recent years that pre-trained language models are capable of either fine-tuning or solving a wide range of real-world problems. Moreover, performance tends to get better with increasing scale. In keeping with this pattern, hundreds of billions of parameters are frequently seen in modern language models. While the public availability of 100B+ parameter models makes them easier to access, they remain difficult to use for the majority of researchers and practitioners due to memory and computational costs. Distributed computing techniques allow multiple users to collaborate and perform inference and fine-tuning of large language models over the Internet. Each participant runs a server, a client, or both. A server hosts a subset of model layers (typically, Transformer blocks) and handles requests from clients. A client can form a chain of pipeline-parallel consecutive servers to run the inference of the entire model.

The objective of employing Distributed Computing Techniques in the training and inference of Large Language Models (LLMs) is to enhance the efficiency, scalability, and speed of the model's learning and application processes \cite{DBLP:journals/corr/abs-2312-12705}. Distributed computing involves the use of multiple interconnected processors or nodes working collaboratively to perform computational tasks. In the context of LLMs, this approach is adopted to address the challenges associated with the sheer size and complexity of these models, allowing for parallelized training and accelerated inference.

% \chapter{A}
% % @Sayem -- add as needs be  

% \chapter{B}

\section{\textsc{PETALS Metaheuristic Upgrade: Finding the Optimal Chain}}
% @Sayem 
In this section, we introduce PETALS, a novel system that aims to transform the large language model (LLM) inference and fine-tuning process. PETALS addresses the need for cost-effective LLM utilization by pooling resources collaboratively. We discuss the core objectives and features that set the foundation for our proposed approach. Moving on to the next section, we present our metaheuristics strategy for finding the optimal chain for selecting a sequence of servers capable of executing the required layer.
\subsection{Background}
\subsubsection{PETALS}
As Natural Language Processing (NLP) tasks increasingly demand the utilization of Large Language Models (LLMs) with over 100 billion parameters, accessibility to these models becomes paramount for researchers and developers. While recent releases such as BLOOM-176B\cite{le2023bloom} and OPT-175B\cite{zhang2022opt} offer pre-trained models of unprecedented scale, the computational requirements for utilizing these models pose a significant barrier, often exceeding the capabilities of standard hardware setups. 

Addressing this critical issue, the recent paper on PETALS\cite{borzunov2022petals} introduces a novel system tailored for the inference and fine-tuning of large language models collaboratively by pooling together the resources of multiple parties. This innovative approach seeks to overcome the limitations of traditional methods of model offloading, offering an efficient solution for running inference tasks on consumer-grade GPUs, achieving approximately one step per second for models like BLOOM-176B. Such performance levels are crucial for enabling interactive applications reliant on large language models, significantly enhancing accessibility and usability.

\subsubsection{Distributed Inference and Fine-tuning of LLMs Over The Internet}
Distributed Inference and Fine-tuning of Large Language Models Over The Internet\cite{petals2} paper investigates methods for cost-efficient inference and fine-tuning of Large Language Models (LLMs), comparing local and distributed strategies. The authors demonstrate that models exceeding 50 billion parameters can run efficiently on geo-distributed devices within a consumer-grade network. Addressing challenges such as device disconnection and uneven hardware, the authors develop fault-tolerant inference algorithms and load-balancing protocols showcased in PETALS\cite{borzunov2022petals}. These innovations maximize system throughput, contributing to the accessibility and efficiency of LLMs.

\subsubsection{PETALS' Optimal Chain Strategy}
 In Algorithm 1 (Generating sequence) of Distributed Inference and Fine-tuning of Large Language Models Over The Internet\cite{petals2} paper, the find\_best\_chain function selects a sequence of servers that can run the required layers in the least amount of time. To estimate this time they add up two factors: computation time, determined by the server’s compute throughput (“GPU speed”), and the network latency between the client and that server. Servers measure their own compute throughput and share this information with the clients. Clients measure the network latency between them and a given server by “pinging” the candidate servers during routing. To find the best chain of servers, clients find the shortest path between the first and last block, using a graph where edge weights correspond to server inference time.

\subsection{Our Proposed Approach}
In this project, we have introduced a metaheuristics approach to find the optimal chain for selecting a sequence of servers capable of executing the required layers. Our optimization process considers two primary objectives: the compute throughput of the servers (referred to as "GPU speed") and the network latency between the client and each server. To address this multi-objective optimization problem effectively, we have implemented the Non-Dominated Sorting Genetic Algorithm II (NSGA-II)\cite{NSGA-ii-2002fast}.

\subsubsection{NSGA-II Algorithm}
% \textbf{\textcolor{blue}{Write about NSGA-II}}
NSGA-II\cite{NSGA-ii-2002fast}, which stands for Non-Dominated Sorting Genetic Algorithm II, is a powerful and widely used multi-objective optimization algorithm. It is designed to address problems with multiple conflicting objectives, where finding a single optimal solution is not feasible due to the trade-offs between objectives. NSGA-II efficiently explores the solution space to identify a set of solutions known as the Pareto front, representing the best trade-offs between conflicting objectives.

NSGA-II efficiently explores and identifies the Pareto-optimal solutions, i.e., the solutions that are not dominated by any other solution in terms of both objectives. By employing NSGA-II, we aim to discover a set of chains that represent the best trade-offs between GPU speed and network latency, providing a range of options for selecting the most suitable chain according to specific requirements and constraints.

\subsubsection{Initial Population}
NSGA-II will maintain a population of size m (m=100 in our implementation). Each individual in the population represents a candidate solution. We will employ a matrix-based representation to model a candidate solution for our server chain selection problem. In this representation:
\begin{itemize}
    \item Each row of the matrix corresponds to a server.
    \item Each column of the matrix represents a block, indicating whether the server is utilized for training/inference in that particular block.
    \item The values in the matrix indicate whether the corresponding server is used in that particular block.    
\end{itemize}

\subsubsection{Evaluation}
 NSGA-II will evaluate each candidate solution by calculating its fitness with respect to the multiple objectives. In our case, the objectives are the server's compute throughput (GPU speed) and the network latency between the client and that server.

\subsubsection{Our Implementation}
We have utilized the pymoo\cite{2020pymoo} library, a Python-based framework for multi-objective optimization, specifically implementing the NSGA-II algorithm.

In order to implement our approach, we first needed to define the problem using the pymoo library. To achieve this, we inherited the Problem class from the pymoo library and created a new class called ChainSequence, which defined the problem. Our objective was to minimize latency and maximize throughput, which was achieved through the evaluate function of the ChainSequence class. The objective function aimed to minimize the sum of latencies and maximize the sum of throughputs across all blocks. We also defined a constraint in the evaluate function, which stipulated that each block must be assigned to at least one server.

PETALS has two modes for creating chain sequences: one for minimum latency and one for maximum throughput.

We have implemented a new mode called "Latency-Throughput-Tradeoff" to create an optimal chain using our approach. To achieve this, we have created an instance of the ChainSequence class and utilized Bit-flip mutation and Single Point Crossover as operators. We have used the NSGA-II algorithm to find the best chain.

\subsubsection{Experimental Results}
To conduct a comparison experiment, we need to run PETALS as a client. However, PETALS does not provide the option to run as a client in the public swarm, making it impossible to test our approach. Although we can connect to PETALS as a server and train our models, we require PETALS to run as a client to test our approach. Unfortunately, PETALS restricts modified code from running on the public swarm, which means we need a private swarm to test our approach. Due to a lack of resources, we are unable to conduct the comparison experiment. If we had sufficient resources and launched a private swarm, we would be able to conduct the comparison experiment.

\subsubsection{Code Availability}
The code can be found \href{https://github.com/SaemHasan/petals-main/tree/main}{here}. You can find our modifications and proposed approach in the \href{https://github.com/SaemHasan/petals-main/blob/main/src/petals/client/routing/sequence_manager.py}{sequence manager} file.

\section{Distributed LLM Serving Techniques}

A serving system is responsible for managing endpoints that receive inference requests, scheduling the execution of an execution engine, and sending responses to these requests. An execution engine, on the other hand, is tasked with driving the main mathematical operations. In this chapter, we present a comparative study on three of the state-of-the-art LLM serving systems and explore how they employ distributed computing techniques. 

\subsection{Background}

\subsubsection{Autoregressive Nature of Generative LLMs}

Generative Large Language Models (LLMs), such as GPT (Generative Pre-trained Transformer), possess an autoregressive nature, where they generate sequences of tokens one step at a time. This means that during the generation process, each token is produced based on the previous tokens in the sequence. By leveraging this autoregressive mechanism, LLMs can capture complex patterns and dependencies within the data they were trained on, allowing them to generate coherent and contextually relevant text.

\begin{figure}[h]
    \centering
    \includegraphics[width=0.4\textwidth]{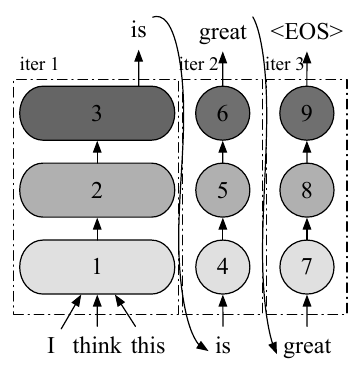}
    \caption{A computation graph representing the inference procedure of a GPT-like model}
    \label{fig:gpt_ar}
\end{figure}

As we see in Figure \ref{fig:gpt_ar}, \verb|I think this| is the initial input to the model. In \textbf{one} iteration, the model passes through \textbf{all} of its layers to generate the next token \verb|is|. This token is also treated as input for the next iteration. The model again makes a pass over all of its layers to generate the next token \verb|great|. The same process is repeated till the special end-of-sentence token is generated.

\subsubsection{The Transformer} 

In the context of generative Large Language Models (LLMs) such as GPT, attention mechanisms are crucial for focusing on relevant parts of the input text during the generation of output sequences. These mechanisms rely on three main learnable components: query, key, and value. The query represents the current token being generated, while keys and values represent different positions or tokens in the input sequence. By comparing the query with keys, the model determines the relevance of each part of the input sequence and assigns appropriate attention weights to the corresponding values. This allows the model to capture long-range dependencies and generate coherent, contextually relevant text by considering the entire input context rather than just adjacent tokens.

\begin{figure}[h]
    \centering
    \includegraphics[width=0.35\textwidth]{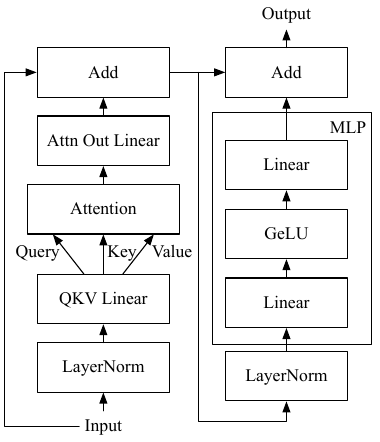}
    \caption{A transformer layer used in GPT}
    \label{fig:enter-label}
\end{figure}

As Attention relies on the keys and values of preceding tokens, we view these keys and values as internal states that must be preserved throughout multiple iterations. A straightforward, state-less inference process would involve recalculating all keys and values for every iteration, considering all tokens in the sequence, including both the client-provided input tokens and the output tokens generated thus far. To prevent such redundant computation,  incremental decoding has been proposed \cite{fairseq}, which preserves the keys and values for reuse in subsequent iterations.

\subsection{ORCA} 

In this subsection, we will review ORCA: A Distributed Serving System for Transformer-Based Generative Models \cite{orca}. 

\subsubsection{Challenges and Solutions}

\subsubsubsection{C1: Early-finished and late-joining requests}

The authors of ORCA note that current systems suffer from a limitation where the serving system and execution engine only coordinate when scheduling the next batch or when the current batch is completed. This approach is problematic for generative models, as requests within a batch may finish at different times due to the varying iterations needed. For example, if one request finishes early while another is still processing, the engine still computes for both, leading to inefficiency.  Additionally, finished results are only returned to the serving system after the entire batch is processed, causing extra latency. 

To illustrate the above, let us consider two prompts: \verb|The sky is very| and \verb|Our society| \verb|exists to remind us that|. The first answer is expected to end within a word, e.g. \verb|cloudy|. The second prompt will result in a much longer sentence. Both latency and throughput will suffer. 

This issue doesn't occur during language model training, as the entire batch is processed in one iteration.

\subsubsubsection{Sol1: Iteration Based Scheduling}

\begin{figure}[h]
    \centering
    \includegraphics[width=\linewidth]{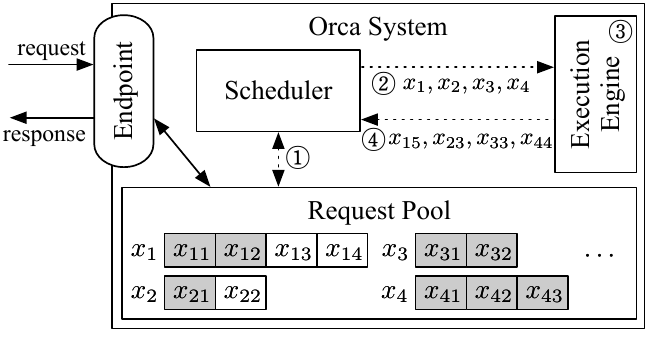}
    \caption{System overview of ORCA}
    \label{fig:orcha_arch}
\end{figure}

The authors propose scheduling executions based on \textbf{iterations} rather than \textbf{batches}. In this approach, the scheduler operates as follows: first, it selects requests for the next iteration; then, it instructs the engine to execute one iteration for the chosen requests; finally, it receives the results of the executed iteration. By receiving feedback on each iteration, the scheduler can detect when a request is completed and promptly deliver its generated tokens to the client. Additionally, when a new request arrives, it can begin processing immediately after the current iteration, reducing queueing delay significantly. With iteration-level scheduling, the scheduler has full control over the number and selection of requests processed in each iteration. The process is depicted in Figure \ref{fig:orcha_arch}.

\subsubsubsection{C2: Batching an arbitrary set of requests}

When attempting to implement iteration-level scheduling in practice, a significant challenge arises concerning batching. To ensure high efficiency, the execution engine must be capable of processing any selected set of requests in a batched manner. Without batching, processing each selected request individually would negate the substantial parallel computation capabilities of GPUs. However, batching can only be applied if the two selected requests are in the same phase, either with the same number of input tokens for the initiation phase or with the same token index for the increment phase. Therefore, the proposed iteration-level scheduling will yield very low throughput.  

\subsubsubsection{Sol2: Selective Batching}

The authors note that not all operations require strict compatibility with batch size. While attention-related layers would require tensors of the same shape to be batched, other layers (linear, layer normalization, activation, addition) can be executed by flattening the input matrices without any compatibility issue.

\subsubsection{Architecture} 

\begin{figure}[h]
    \centering
    \includegraphics[width=\linewidth]{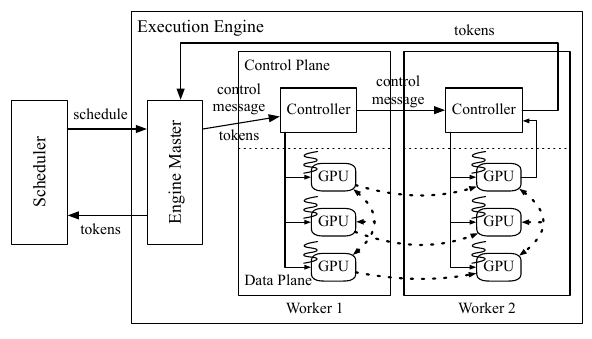}
    \caption{An illustration of the distributed architecture of ORCA’s execution engine}
    \label{fig:orca_dist}
\end{figure}

ORCA employs known parallelization techniques for Transformer models, including intra-layer and inter-layer parallelism, originally developed for distributed training. Intra-layer parallelism splits matrix multiplications (e.g., Linear and Attention operations) and their parameters across multiple GPUs, while inter-layer parallelism splits Transformer layers over GPUs. The ORCA execution engine supports distributed execution, with each worker process managing an inter-layer partition of the model and coordinating GPU control. During execution, the engine forwards batch information to worker processes, which parse and issue GPU kernels without waiting for completion, minimizing CPU-GPU synchronization. Unlike existing systems, ORCA separates communication channels for control messages and tensor data transfer, reducing overhead by avoiding GPU-to-GPU communication for CPU-issued control messages. An illustration is shown in Figure \ref{fig:orca_dist}.

\subsection{vLLM}

We now review the concepts of vLLM and PagedAttention from the paper ``Efficient Memory Management for Large Language
Model Serving with PagedAttention" \cite{vllm}. 

\begin{figure}
    \centering
    \includegraphics[width=\linewidth]{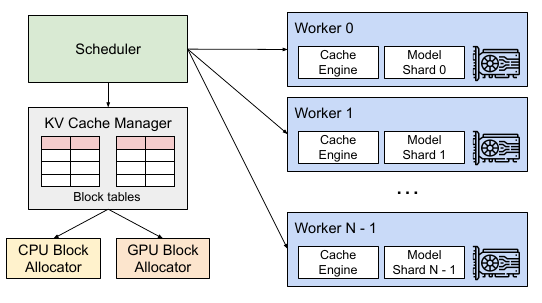}
    \caption{An overview of vLLM}
    \label{fig:enter-label}
\end{figure}

\subsubsection{Challenges}

The transformers need to reuse their keys and values, referred to as KV cache, by the authors. The authors note that the existing systems, including ORCA, suffer from internal and external memory fragmentation. To ensure contiguous storage of the KV cache for a request, they allocate a contiguous memory chunk based on the request's maximum length (e.g., 2048 tokens). However, this can lead to severe internal fragmentation, as the actual request length may be much shorter. Additionally, even if the actual length is known beforehand, pre-allocation is inefficient because the entire chunk remains reserved during the request's lifetime, preventing other shorter requests from utilizing unused parts. Moreover, external memory fragmentation can also occur since the pre-allocated size may vary for each request. In the profiling experiments performed by the authors, ORCA utilizes only 20.4\% - 38.2\% of the KV cache memory to store actual token states.

Furthermore, current systems fail to capitalize on memory-sharing opportunities. Advanced decoding algorithms used in LLM services, such as parallel sampling and beam search, generate multiple outputs per request, potentially allowing sequences within the request to share their KV cache partially. However, existing systems do not enable memory sharing since the KV cache of sequences is stored in separate contiguous spaces.

The authors propose \textit{PagedAttention}, an attention algorithm that
operates on KV cache stored in non-contiguous paged
memory, which is inspired by the virtual memory and
paging in OS. Then, they design and implement \textit{vLLM}, a distributed LLM serving engine built on top of \textit{PagedAttention}. 

\subsubsection{PagedAttention}

Paging in operating systems is a memory management scheme that allows the physical memory to be divided into fixed-size blocks called pages, which are then used to store and manage virtual memory addresses. Pages are transferred in and out by the OS, based on the requests from processes. The authors suggest that one can think of blocks of KV cache as pages, LLM's tokens as bytes, and requests as processes. 

The usual attention mechanism computes a weighted sum of the values \(V\) based on the similarity between the query \(Q\) and the keys \(K\). The attention formula used in the Transformer model is given by:

\[
\text{Attention}(Q, K, V) = \text{softmax}\left(\frac{QK^T}{\sqrt{d_k}}\right)V
\]

Here, \(Q\), \(K\), and \(V\) represent the query, key, and value matrices, respectively. The term \(d_k\) represents the dimensionality of the keys and queries.

Instead of performing the matrix multiplications at a time, which would require the tensors to be stored in contiguous memory, the authors utilize the concept of block matrix multiplication. It involves breaking large matrices into smaller blocks, performing matrix multiplication on these blocks, and summing the results to obtain the final output. This approach turns out beneficial for handling large matrices efficiently, enabling parallelization, and reducing memory requirements. In this specific case, the KV cache is divided into blocks of equal length. Only the blocks those are needed are paged in for computation. 

\subsubsection{KV Cache Manager}

\begin{figure}[h]
    \centering
    \includegraphics[width=\linewidth]{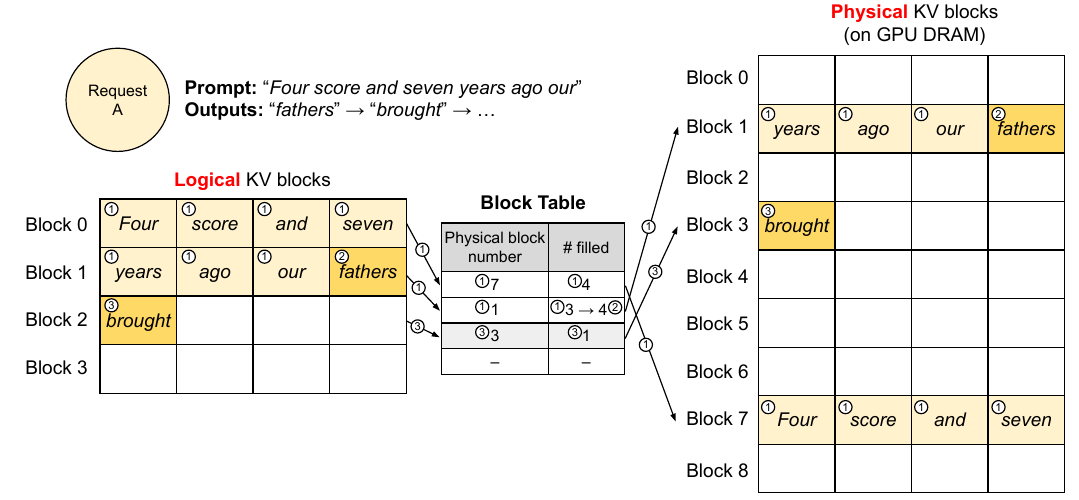}
    \caption{Block table translation in vLLM.}
    \label{fig:vllm_blocks}
\end{figure}

vLLM's memory management system draws parallels with virtual memory in operating systems, where memory is partitioned into fixed-sized pages, enabling logical pages of user programs to map to non-contiguous physical pages. Similarly, vLLM organizes the KV cache into fixed-size blocks akin to virtual memory pages, dynamically allocating physical pages as needed. This approach eliminates memory waste, as vLLM can grow the KV cache memory dynamically without reserving it in advance, improving efficiency compared to existing systems. An illustration is shown in Figure \ref{fig:vllm_blocks}. 

\subsubsection{Distributed Execution}

vLLM employs a model parallelism approach to handle the vast number of parameters in Large Language Models. This strategy follows an SPMD execution schedule, where linear layers are partitioned for block-wise matrix multiplication, and GPUs synchronize intermediate results via an all-reduce operation. Despite model parallel execution, each shard processes the same input tokens, necessitating a single KV cache manager within the centralized scheduler. GPU workers share this manager and the mapping from logical to physical blocks, allowing them to execute the model efficiently with the provided physical blocks for each input request. Although each GPU worker has the same physical block IDs, they store only a portion of the KV cache for their corresponding attention heads.
\subsection{InfiniteLLM} 

We now review the manuscript ``\textit{Infinite-LLM}: Efficient LLM Service for Long Context with \textit{DistAttention} and \textit{Distributed KVCache}" \cite{infllm}.  

\subsubsection{Issues with PagedAttention}

The authors identify three main issues with \textit{PagedAttention} and try to overcome them.

\begin{itemize}
    \item The memory swapping scope of Page-dAttention was confined to GPU and CPU memory within a single node, limiting its ability to handle extremely long context lengths.
    \item While its paging strategy aimed to reduce memory fragmentation, it swapped entire KV Caches at a request-level, missing the opportunity for more adaptive and granular scheduling in a distributed cloud environment.
    \item Interruptions in computation due to swapped-out requests can lead to inconsistent performance for running tasks, potentially violating strict service-level agreements.
\end{itemize}

\subsubsection{Improvements}

The proposed \textit{DistAttention} mechanism disaggregates the KV cache into smaller sub-blocks, enabling distributed memory management. This algorithm partitions into multiple Micro Attentions (MAs), each handling a subset of KV cache tokens independently, and aggregates their results for the final attention computation. The DistKV-LLM aims to streamline KV Cache management, coordinating memory usage across GPUs and CPUs in the data center. In cases of memory shortage, it proactively reallocates available memory from instances with excess capacity. This automated process involves two main components: the rManager, which virtualizes GPU and CPU memories within each LLM service instance, handling memory requests locally and remotely, and the gManager, a global coordinator ensuring coherent resource management among distributed rManagers, promoting effectiveness and scalability. 

\subsubsection{rManager and gManager}

\begin{figure}
    \centering
    \includegraphics[width=\linewidth]{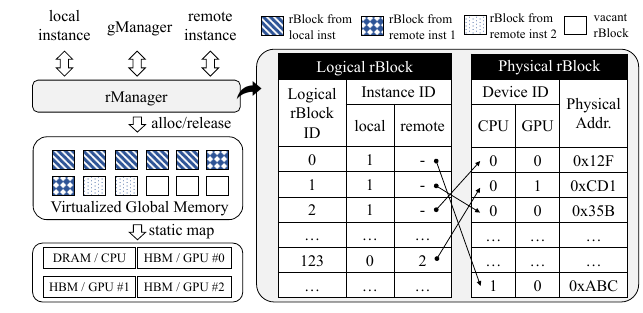}
    \caption{Illustration of the rManager design}
    \label{fig:rManager}
\end{figure}

 The Key-Value (KV) caches in Large Language Models (LLMs) are divided into smaller units called rBlocks, each containing a fixed number of Key-Value tokens and associated metadata. This metadata includes information about the rBlock's ID, Instance ID, device ID, and physical ID, indicating its location within the memory hierarchy. Each LLM service instance features a dedicated rBlock manager, the rManager, responsible for managing rBlocks on local devices and virtualizing GPU memory space. The rManager maintains a mapping table linking logical rBlocks to their corresponding physical addresses in global memory, as shown in Figure \ref{fig:rManager}. 

The main distributed computing component of \textit{InfiniteLLM} is its \textit{gManager}, which maintains global memory information across all instances by collecting periodic heartbeat signals. Using this data, the gManager constructs a detailed global debt ledger, tracking available memory space. When an instance requires additional memory for rBlocks, its rManager acts as a debtor, querying the gManager for potential creditor instance IDs. The gManager selects potential creditors based on locality, availability, and communication cost, proposing three recommendations. The debtor instance sequentially requests memory from these creditors until successful allocation, ensuring efficient and responsive memory space management across the data center.

\begin{figure}
    \centering
    \includegraphics[width=\linewidth]{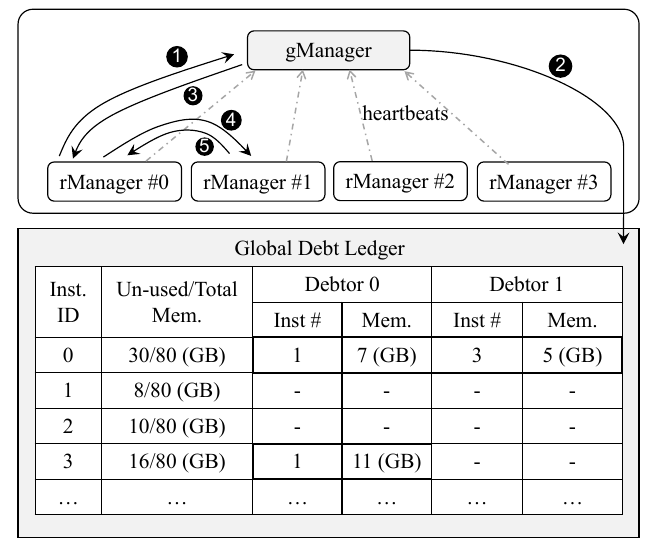}
    \caption{Illustration of gManager and global debt ledger.}
    \label{fig:gManager}
\end{figure}

An illustration is shown in Figure \ref{fig:gManager}. The table contains information regarding available memory spaces and spaces lent to debtors, depicting the memory usage dynamics of five instances. Instance 0, with a light workload, lends spaces to instances 1 and 3. Instance 1, managing a lengthy context, borrows space from both instances 0 and 3. Instance 2 neither borrows nor lends. Instance 3 finds itself in a dual role, borrowing from instance 0 while lending to instance 1 simultaneously.

\subsection{Performance Comparison}

In this subsection, we present the comparison among the three serving frameworks presented above. Since these serving systems and their underlying language models are huge, it was beyond our computing abilities to perform the ablation studies ourselves. We tabulate the results, as reported by the respective authors. 

\subsubsection{ORCA vs vLLM}

The authors of vLLM compare their model with two baselines: FasterTransformer \cite{li-etal-2022-easy} and three versions of ORCA based on varying levels of space over-reservation for request outputs: Orca (Oracle) assumes perfect knowledge of output lengths, Orca (Pow2) over-reserves by up to 2x, and Orca (Max) always reserves space up to the maximum sequence length of 2048 tokens. They used OPT models of several sizes as the underlying LLM \cite{zhang2022opt}. The datasets used for inference were Alpaca \cite{alpaca} and ShareGPT \cite{ShareGPT}. The metric they consider is \textit{normalized latency},  the mean of every request’s end-to-end latency divided by its output length.

%\singlecolumn 
\begin{figure*}
    \centering
    \includegraphics[width=\textwidth]{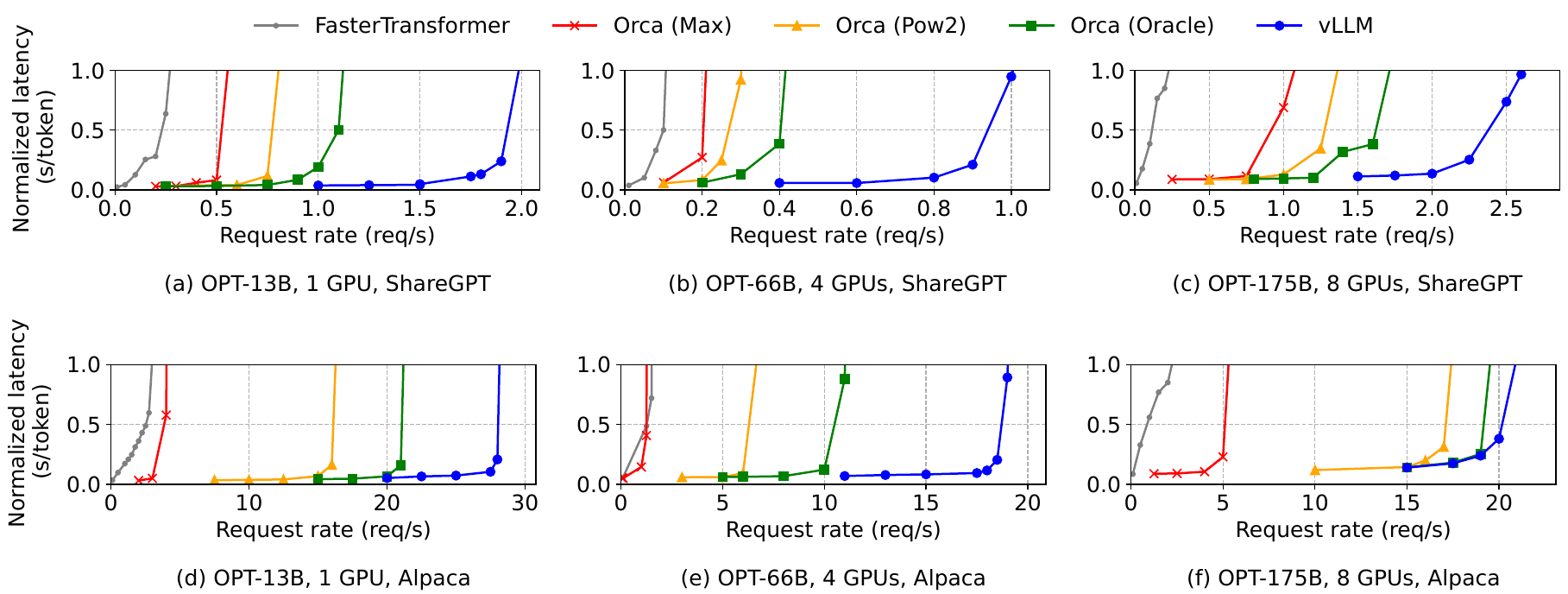}
    \caption{Single sequence generation with OPT}
    \label{fig:orca_vllm}
\end{figure*}
%\twocolumn 
The results are presented in Figure \ref{fig:orca_vllm}. On the ShareGPT dataset, vLLM can sustain 1.7x–2.7x
higher request rates compared to Orca (Oracle) and 2.7x–8x
compared to Orca (Max), while maintaining similar latencies. The result on the Alpaca dataset follows a similar trend. 

\subsubsection{vLLM vs DistKV-LLM}

\begin{figure*}
    \centering
    \includegraphics[width=\textwidth]{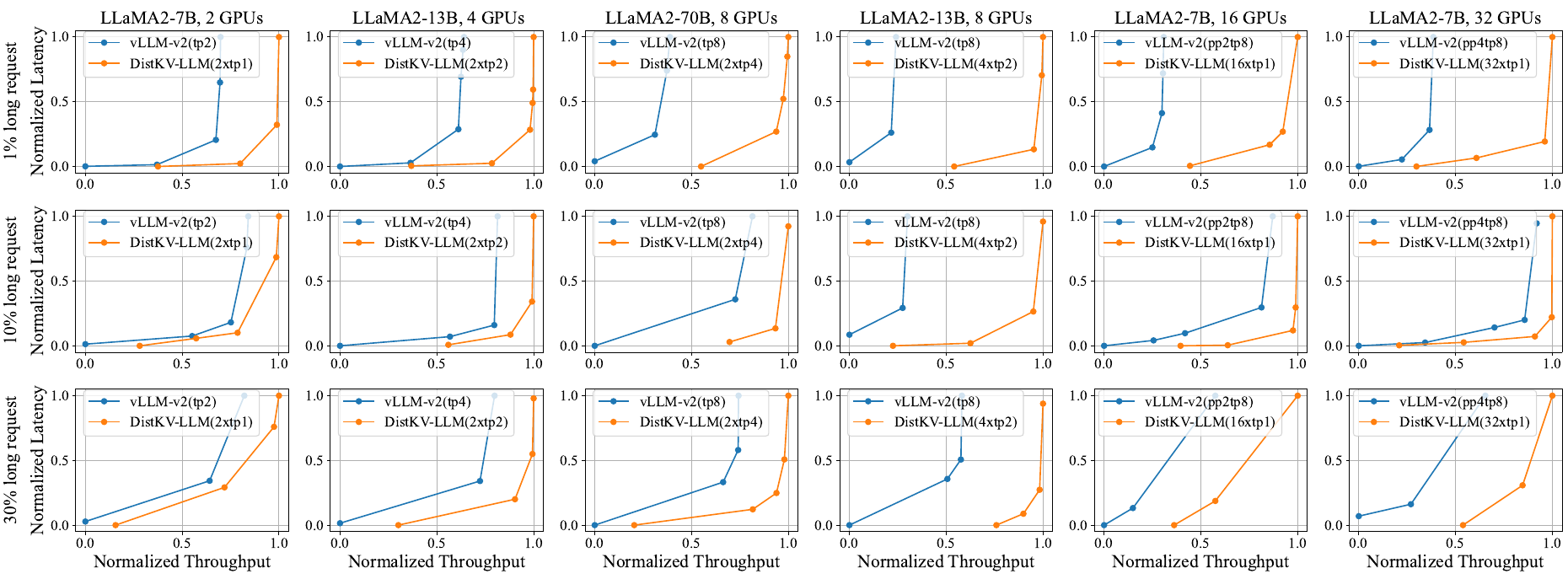}
    \caption{End-to-end serving performance of vLLM and InfiniteLLM}
    \label{fig:vllm_inf_llm}
\end{figure*}

The authors of DistKV-LLM evaluated their performance using datasets with varying proportions of long requests. Results, depicted in Figure \ref{fig:vllm_inf_llm}, show that DistKV-LLM improves throughput by 1.4x to 2.4x over the baseline with 1\% long requests, and by 1.1x to 1.4x with 10\% long requests. However, as the proportion of long requests increases, the performance gain diminishes due to reduced advantage in linear computations and comparable performance in attention computations compared to the baseline.

\section{Conclusion}

Our investigation into distributed computing techniques for large language models (LLMs) underscores their pivotal role in realizing the full capabilities of these powerful AI systems. We first discussed how LLMs can be democratized using distributed computing techniques. One limitation of this approach we note is that \textsc{Petals} runs as a public swarm where anyone can contribute their computing powers. However, there is no reward system inherent in \textsc{Petals}. There are some distributed systems where contribution to the system creates some extant value, for example, in blockchains \cite{bitcoin, ethereum}. As for LLM serving, we have seen that our three state-of-the-art methods each start with some algorithmic improvements that make LLMs more suitable to be distributed across several worker nodes. As highlighted by \cite{mi}, exploring alternative architectures and/or implementing efficient decoding algorithms considering distributed computing principles is imperative for ensuring the proper scalability of future large language models.
% @Fahim 

%\chapter{Fault Tolerance in LLMs}

%\chapter{Consensus with LLMs} 

%\section{Conclusion}

% \section*{References}
\bibliographystyle{abbrv}
\bibliography{ref}

\end{document}